# Mixbiotic society measures: Comparison of organizational structures based on communication simulation

Short title: mixbiotic society measures

Takeshi Kato[1*], Jyunichi Miyakoshi[1], Tadayuki Matsumura[2], Yasuyuki Kudo[1], Ryuji Mine[1], Hiroyuki Mizuno[2], Yasuo Deguchi[3]

[1] Hitachi Kyoto University Laboratory, Open Innovation Institute, Kyoto University, Kyoto, Japan

[2] Hitachi Kyoto University Laboratory, Center for Exploratory Research, Research & Development Group, Hitachi, Ltd., Tokyo, Japan

[3] Department of Philosophy, Graduate School of Letters, Kyoto University, Kyoto, Japan

\* Corresponding author

E-mail: kato.takeshi.3u@kyoto-u.ac.jp (TK)

# Abstract

The philosophical world has proposed the concept of "mixbiotic society," in which individuals with freedom and diverse values mix and mingle to recognize their respective "fundamental incapability" each other and sublimate into solidarity, toward solving the issues of social isolation and fragmentation. Based on this concept, the mixbiotic society measures have been proposed to evaluate dynamic communication patterns with reference to classification in cellular automata and particle reaction–diffusion that simulate living phenomena. In this paper, we applied these measures to five typologies of organizational structure (Red: impulsive, Amber: adaptive, Orange: achievement, Green: pluralistic, and Teal: evolutionary) and evaluated their features. Specifically, we formed star, tree, tree+jumpers, tree+more jumpers, and small-world type networks corresponding to each of five typologies, conducted communication simulations on these networks, and calculated values for mixbiotic society measures. The results showed that Teal organization has the highest value of the mixism measure among mixbiotic society measures, i.e., it balances similarity (mixing) and dissimilarity (mingling) in communication, and is living and mixbiotic between order and chaos. Measures other than mixism showed that in Teal organization, information is not concentrated in a central leader and that communication takes place among various members. This evaluation of organizational structures shows that the mixbiotic society measures is also useful for assessing organizational change. In the future, these measures will be used not only in business organizations, but also in digital democratic organizations and platform cooperatives in conjunction with information technology.

# Introduction

The philosopher Deguchi proposes the concept of "mixbiotic society" that takes the symbiotic society a step further, toward solving the issues of social isolation and fragmentation [1–3]. Social isolation and fragmentation are generally social issues related to communication [4–7], which are viewed as problems of community impoverishment (individual atomization: atomism) and hypertrophy (in-group crowding: mobism), respectively [8, 9].

The "mixbiotic society" is a society in which individuals with freedom and diverse values mix and mingle in physical proximity to recognize their respective "fundamental incapability" each other and sublimate into solidarity. The "fundamental incapability" is that the individual "I" is incapable of any physical action alone, nor of complete control over others. The subject of mixbiotic society is not "Self as I," but "Self as WE" who are entrusted to each other. And "Self as WE" requires both openness (freedom) to avoid fragmentation and fellowship (solidarity) to avoid isolation. Deguchi also proposes "well-going" as a dynamic life as opposed to well-being as a static way of being.

The "Mixbiotic Society Measures (MSM)" have been proposed to evaluate mixbiotic societies [10]. The MSM assesses the dynamic patterns of communication generation and disappearance on communication networks, based on the theory that social systems are autopoietic systems consisting of recursive process networks of communication generation and disappearance [11, 12]. Compared to social network analysis [13] and temporal network analysis [14], which are based on the analysis of static network structures, MSM is characterized by evaluation of the relative change in communication patterns and their variance on the network with reference to classification in cellular automata (CA) and particle reaction–diffusion (PRD) [15, 16], which simulate living phenomena. As MSM corresponding to four phases of society, a mixism measure which indicates well-going of a desirable mixbiotic society, and a nihilism measure which indicates no communication, are defined, along with an atomism measure which indicates isolation, and a mobism measure which indicates fragmentation.

In the above literature [10], along with the definition of MSM, the usefulness of mixism measure and the possibility of typifying communities with multiple measures are presented through validation with real-society datasets (high school, elementary school, workplace, village, conference, online community, email). However, this literature has yet to examine the organizational structure of workplaces and business cooperation. Along with communities that support human life, organizations that support economic activity have been an important concern of sociology since olden times (e.g., [17, 18]). This is because organizational structure defines the interactions of its constituent members and influences goal achievement and persistence in the organization (see review articles [19, 20]).

Five typologies (Red, Amber, Orange, Green, and Teal) presented by organizational change consultant Laloux are well known as organizational structures [21, 22]. Briefly introduced, a Red organization is an impulse organization with a leader with strong power and all other members uniformly aligned around leader, as metaphorized by a pack of wolves. An Amber organization is an adaptive organization with a hierarchical and rigid chain of command, as in a military metaphor. An Orange organization is an achievement organization that are fundamentally hierarchical, but with connections among members that transcend hierarchy, as in a machine metaphor. A Green organization is a pluralistic organization in which hierarchy still remains, but the connections among members are further developed, as in a family metaphor. A Teal organization is an evolutionary organization in which power hierarchies disappear, but they are not merely flat; authority is decentralized and roles are naturally differentiated, as in a living organism. Teal organizations are living organizations and related to mixbiotic societies.

Therefore, the purpose of this paper is to newly evaluate and characterize Laloux's five typologies by MSM with respect to organizational structures that have not yet been examined in the literature [10]. Specifically, after forming networks corresponding to five typologies, we simulate the generation and disappearance of communication on the networks based on the communication models in the literature [10], and evaluate the temporal changes in these patterns by MSM. The remainder of this paper is organized as follows. In the Methods section, we describe the characteristics of networks corresponding to five typologies, followed by a brief introduction to the communication model and MSM calculations in the literature [10]. In the Results section, we form networks for five typologies and present the results of MSM calculations based on communication simulations. The Discussion section discusses the usefulness of MSM for organizational change and offers perspectives for the future.

# Methods

## Organizational structure and network

In this section, we first describe the network structure corresponding to Laloux's five typologies, followed by a brief introduction to communication models and MSM calculations based on the literature [10].

Table 1 shows the correspondence between five typologies of organizational structure (Red, Amber, Orange, Green, and Teal) and network structures. These typologies are closely related to network structures, as described in the Introduction section. Star and tree networks can be formed using common tools (e.g., [23, 24]). Jumpers are formed by randomly adding edges to a tree network. The Watts-Strogatz (WS) model [25] or the Barabási-Albert (BA) model [26] are used to form small-world and scale-free networks, respectively. Both models are well known for simulating real-world networks. Since MSM's evaluation of communication simulations in the literature [10] did not show a marked difference between WS model and BA model, we focus on the WS model here.

**Table 1. Network structure corresponding to organization structure.**

|  | Red | Amber | Orange | Green | Teal |
|---|---|---|---|---|---|
| Feature | ・One leader<br>・Even follower | ・Hierarchy<br>・Strict | ・Semi-hierarchy<br>・Additional freedom | ・Soft-hierarchy<br>・Empowerment of front-line | ・Decentralized<br>・Natural differentiation |
| Network structure | Star | Tree | Tree + Jumpers | Tree + More jumpers | Small-world or Scale-free |

## Communication model

The role of communication model is to simply model the process of generation and disappearance of communication. Fig 1 shows a simplified computational flow. For more details, please refer to the literature [10]. In this flow, first, as initial settings, the generation rate $g$ and disappearance rate $d$ of communication, the information unit $u$ of communication, the number of vertices $n_0$ that initially give the information unit $u$, and the number of computation steps $t_{max}$ are set. Then, at each step $t$, a sender and a receiver are randomly selected from among the vertices of the network graph $G$ according to the generation rate $g$, an information unit $u$ is sent from the sender to the receiver, and information is randomly erased from the vertices according to the disappearance rate $d$. Finally, we obtain the information set $Q(t) = \{q_1(t), q_2(t), q_3(t), \cdots, q_n(t)\}$ held by $n$ vertices at each time step $t$.

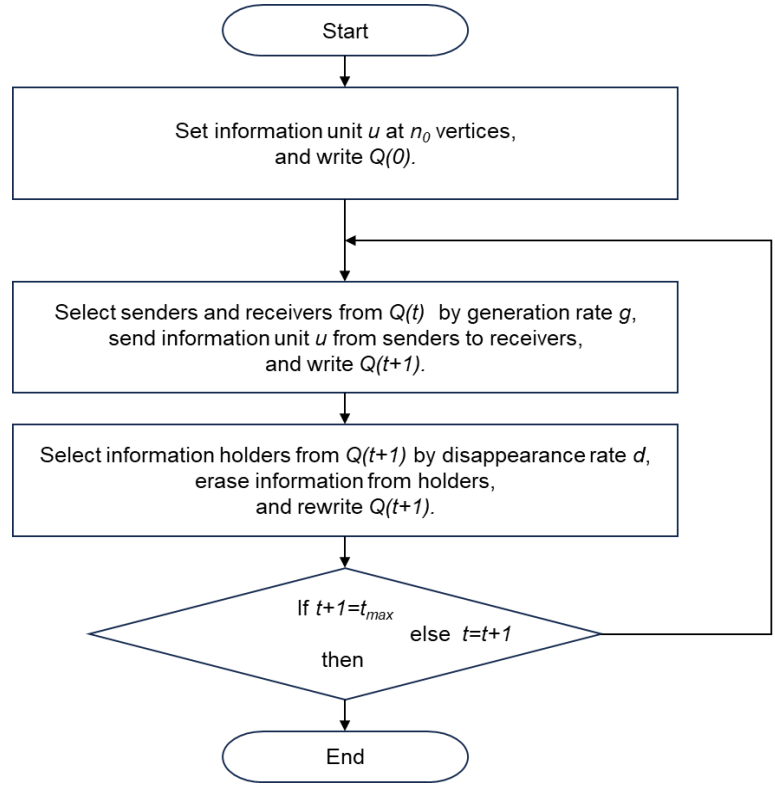

**Fig 1. Computational flow of communication model [10].**

## Mixbiotic society measures

In MSM, the information set $Q$ is regarded as an $n$-dimensional vector as it represents a communication pattern on a network of $n$ vertices, and the change in the information sets $Q(t)$ and $Q(t+1)$ between time $t$ and time $t+1$ is calculated. In the literature [10], the following four quantities of change are listed.

1) Change in total information amount

$$I(t+1) = \frac{|\sum_{i=1}^{n} q_i(t+1) - \sum_{i=1}^{n} q_i(t)|}{n \cdot u}. \tag{1}$$

2) Change in Euclidean distance

$$L(t+1) = \frac{\sqrt{\sum_{i=1}^{n}(q_i(t+1) - q_i(t))^2}}{\sqrt{n} \cdot u} \quad (2)$$

3) Relative change in Euclidean distance

$$L_R(t+1) = \frac{\sqrt{\sum_{i=1}^{n}(q_i(t+1) - q_i(t))^2}}{\sqrt{\sum_{i=1}^{n} q_i(t+1)^2}}. \quad (3)$$

4) Cosine similarity

$$S(t+1) = \frac{\sum_{i=1}^{n} q_i(t+1) \cdot q_i(t)}{\sqrt{\sum_{i=1}^{n} q_i(t+1)^2} \cdot \sqrt{\sum_{i=1}^{n} q_i(t)^2}}. \quad (4)$$

The MSM derives eight measures by calculating the time mean and variance of the above four variables according to Eqs (1)–(4), and together provides another composite measure for the mixism. It is the mean (similarity) multiplied by the variance (dissimilarity) of the cosine similarity, which represents the balance between mixing (similarity) and mingling (dissimilarity) in a mixbiotic society, and between atomism (atomization) and mobism (crowding) of communication. This corresponds to the large variance of entropy change and the AC component of change in the idea that livingness dwells in class 4 of CA and PRD (between order and chaos, at the edge of chaos) [10].

To the atomism measure, the variance of the relative change in Euclidean distance is corresponded, taking into account that entropy is small and the change in mutual information is large in class 2 (sporadic) of CA and PRD. To the mobism measure, the mean of the Euclidean distance is corresponded, considering the fact that entropy and mutual information are large in class 3 (chaotic). Then, in nihilism, the values of all of the mixism, atomism, and mobism measures are assumed to be almost zero. Table 2 summarizes the set of measures of MSM in the literature [10].

**Table 2. Mixbiotic society measures.**

| # | Measure | Formula | Class/phase |
|---|---|---|---|
| 1 | Total information amount | $\mu_I$ | — |
| 2 | | $\sigma_I^2$ | — |
| 3 | Euclidian distance | $M_{mob} = \mu_L$ | Mobism (Chaotic) |
| 4 | | $\sigma_L^2$ | — |
| 5 | Relative change in Euclidean distance | $\mu_{LR}$ | — |
| 6 | | $M_{atom} = \sigma_{LR}^2$ | Atomism (Sporadic and isolate) |
| 7 | Cosine similarity | $\mu_S$ | — |
| 8 | | $\sigma_S^2$ | — |
| 9 | Composite measure | $M_{mix} = \mu_S \cdot \sigma_S^2$ | Mixism (Living and cyclic) |
| 10 | — | $M_{atom,mix,mob} \approx 0$ | Nihilism (Static and silent) |

# Results

## Network

Fig 2 shows the network graphs formed according to Table 1. Fig 2A is a star network corresponding to Red organization among five typologies of organizational structures, and Fig 2B is a tree network corresponding to Amber organization. Fig 2C is a tree network in Fig 2A with additional jumpers (edges) corresponding to Orange organization, and Fig 2D is a tree network in Fig 2A with additional more jumpers corresponding to Green organization. Fig 2E is a small-world network formed using the Watts-Strogatz (WS) model [25], corresponding to Teal organization. The left side of Fig 2E is drawn as a circle and the right side shows the community partition, both of which are the same network. Fig 2F is a flat hypercube network formed for reference, the left and right sides are the same network.

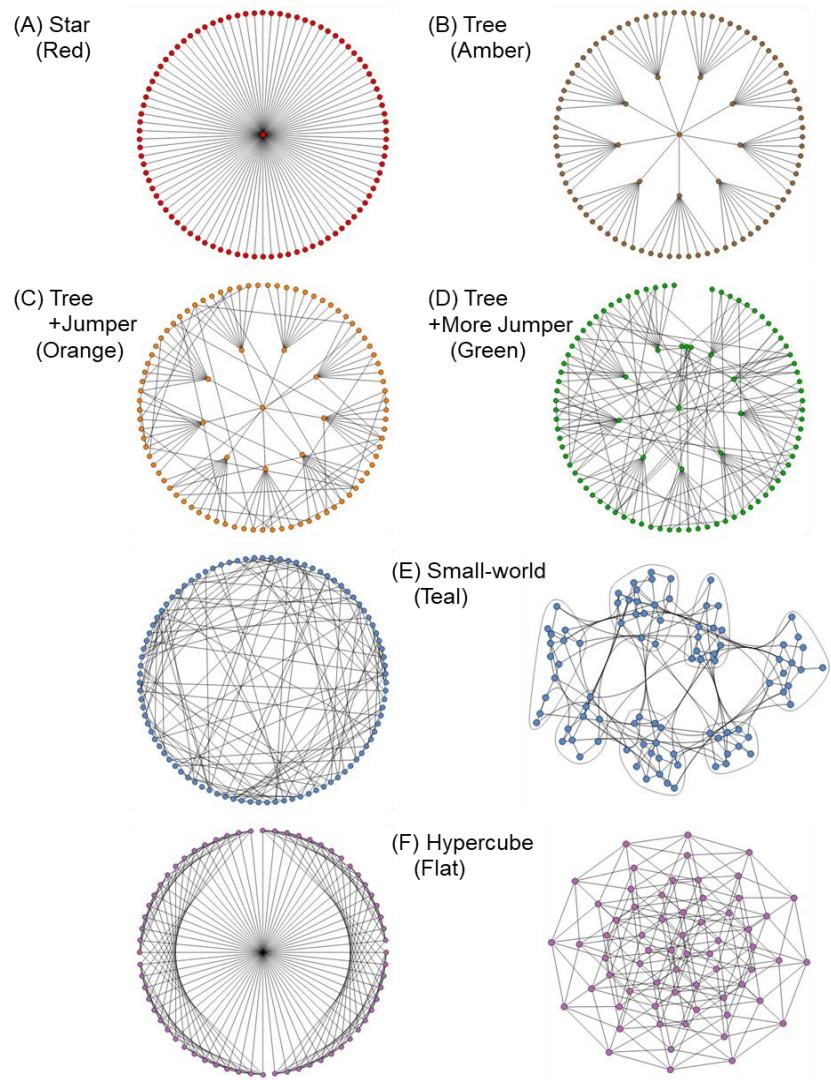

**Fig 2. Network graphs.**

Fig 3 shows the degree distributions of the networks shown in Fig 2. The horizontal axis is the degree, and the vertical axis is the frequency. In the star (Red) in Fig 2A, each vertex has only one edge, except for the central vertex. In the tree (Amber) in Fig 2B, each vertex has only one edge, except for the center vertex and its surrounding vertices. In the tree+jumpers (Orange) in Fig 2C and the tree+more jumpers in Fig 2D (Green), the distribution is more gentle than in the tree (Amber). In small-world (Teal) in Fig 2E, the distribution becomes even smoother due to the elimination of the high degree center vertex. In the hypercube (Flat) in Fig 2F, all vertices have equal degree.

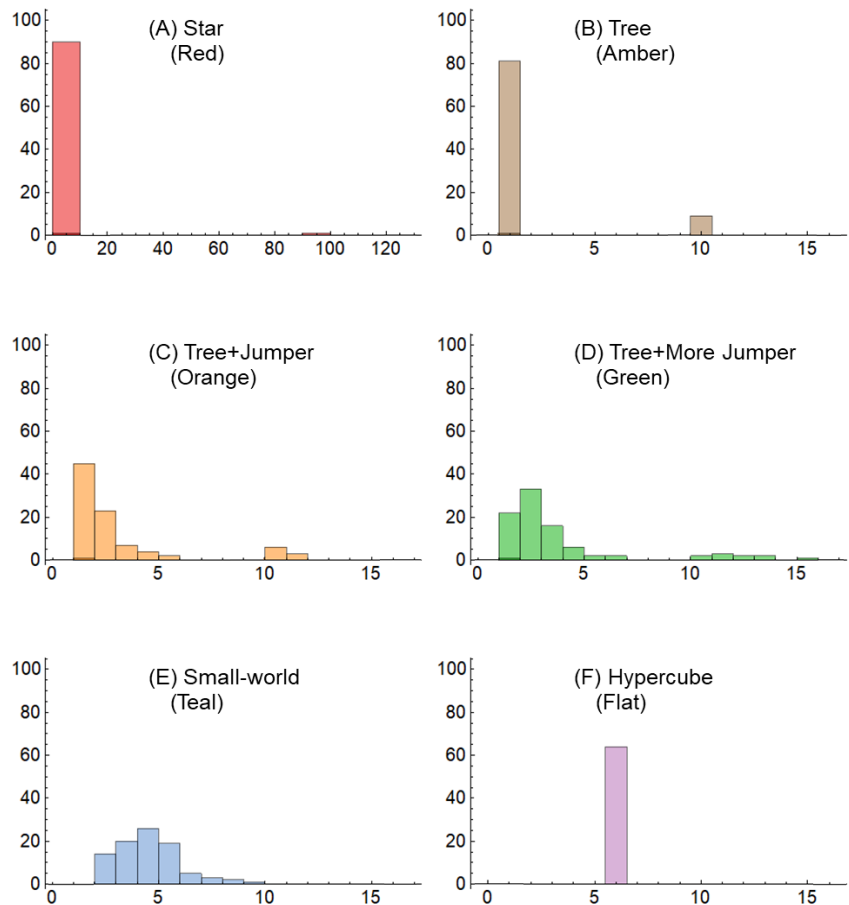

**Fig 3. Degree distributions of network graphs.**

Table 3 shows the graph features of the networks shown in Fig 2. The vertex counts are aligned to the tree network except for the hypercube. Edge count, graph diameter, mean distance, and graph density increase as one goes from star to small-world. In the hypercube, all vertices are evenly connected, the number of edges is large relative to the number of vertices, the mean distance is small, and the graph density is large. The mean cluster coefficient is zero (no clusters) for star and tree, and zero (uniform throughout) for hypercube. In these examples, the tree+more jumpers has the highest cluster coefficient.

**Table 3. Network graph features.**

|  | Star (Red) | Tree (Amber) | Tree +Jumpers (Orange) | Tree+More jumpers (Green) | Small-world (Teal) | Hypercube (Flat) |
|---|---|---|---|---|---|---|
| Vertex count | 91 | 91 | 91 | 91 | 91 | 64 |
| Edge count | 90 | 90 | 120 | 150 | 182 | 192 |
| Diameter | 2 | 4 | 4 | 4 | 6 | 6 |
| Mean distance | 1.98 | 3.56 | 3.41 | 3.21 | 3.38 | 3.05 |
| Density | 0.0220 | 0.0220 | 0.0293 | 0.0366 | 0.0444 | 0.0952 |
| Mean Clustering Coefficient | 0 | 0 | 0.0299 | 0.0665 | 0.0452 | 0 |

# Mixbiotic society measures

The MSM was calculated by simulating communication according to the computational flow shown in Fig 1. The initial settings were set following the literature

[10]: the information unit $u = 1$, the number of vertices $n_0 = 10$ initially given the information unit $u$, and the number of computation steps $t_{max} = 100$. Since atomism, mobism, and nihilism situations are difficult to assume in organizations, four cases of communication generation $g$ and disappearance rate $d$ were set up: $g = 0.4, d = 0.3$, $g = 0.4, d = 0.4$, $g = 0.5, d = 0.4$, and $g = 0.5, d = 0.5$, which the literature [10] has indicated as the range of mixism. Table 4 shows the calcuation results for MSM shown in Table 2 and the network shown in Fig 2. Note that the calculation results were obtained by calculating MSM in one computational flow (the number of steps $t_{max} = 100$) and repeating it 100 times to obtain the average of the number of times.

**Table 4. Calculation results for mixbiotic society measures.**

| Case | Measure | Star (Red) | Tree (Amber) | Tree+Jumpers (Orange) | Tree+More (Green) | Small-world (Teal) | Hypercube (Flat) |
|---|---|---|---|---|---|---|---|
| $g = 0.4$ $d = 0.3$ | $\mu_I$ | 0.0029 | 0.0025 | 0.0047 | 0.0174 | 0.0109 | 0.0241 |
| | $\sigma_I^2$ | 2.96E-04 | 5.22E-05 | 9.79E-05 | 2.69E-04 | 1.21E-04 | 3.58E-04 |
| | $M_{mob} = \mu_L$ | 0.017 | 0.035 | 0.073 | 0.284 | 0.232 | 0.519 |
| | $\sigma_L^2$ | 0.0042 | 0.0071 | 0.0126 | 0.0149 | 0.0120 | 0.0143 |
| | $\mu_{LR}$ | 0.070 | 0.130 | 0.234 | 0.678 | 0.673 | 0.794 |
| | $M_{atom} = \sigma_{LR}^2$ | 0.0590 | 0.1013 | 0.1353 | 0.0921 | 0.1145 | 0.0437 |
| | $\mu_S$ | 0.977 | 0.953 | 0.913 | 0.739 | 0.739 | 0.684 |
| | $\sigma_S^2$ | 0.0080 | 0.0154 | 0.0233 | 0.0246 | 0.0276 | 0.0165 |
| | $M_{mix} = \mu_S \cdot \sigma_S^2$ | 0.0077 | 0.0146 | 0.0205 | 0.0182 | 0.0208 | 0.0113 |
| $g = 0.4$ $d = 0.4$ | $\mu_I$ | 0.0014 | 0.0012 | 0.0016 | 0.0023 | 0.0014 | 0.0103 |
| | $\sigma_I^2$ | 8.62E-05 | 2.68E-05 | 3.16E-05 | 4.68E-05 | 2.34E-05 | 9.29E-05 |
| | $M_{mob} = \mu_L$ | 0.009 | 0.015 | 0.022 | 0.035 | 0.031 | 0.251 |
| | $\sigma_L^2$ | 0.0019 | 0.0029 | 0.0046 | 0.0071 | 0.0060 | 0.0082 |
| | $\mu_{LR}$ | 0.047 | 0.073 | 0.102 | 0.148 | 0.142 | 0.860 |
| | $M_{atom} = \sigma_{LR}^2$ | 0.0461 | 0.0677 | 0.0947 | 0.1216 | 0.1270 | 0.0966 |
| | $\mu_S$ | 0.984 | 0.973 | 0.959 | 0.937 | 0.939 | 0.611 |
| | $\sigma_S^2$ | 0.0059 | 0.0108 | 0.0163 | 0.0249 | 0.0262 | 0.0358 |
| | $M_{mix} = \mu_S \cdot \sigma_S^2$ | 0.0057 | 0.0105 | 0.0156 | 0.0231 | 0.0244 | 0.0223 |
| $g = 0.5$ $d = 0.4$ | $\mu_I$ | 0.0025 | 0.0022 | 0.0052 | 0.0193 | 0.0110 | 0.0237 |
| | $\sigma_I^2$ | 2.91E-04 | 5.11E-05 | 1.23E-04 | 2.99E-04 | 1.21E-04 | 3.46E-04 |
| | $M_{mob} = \mu_L$ | 0.013 | 0.029 | 0.079 | 0.306 | 0.243 | 0.509 |
| | $\sigma_L^2$ | 0.0031 | 0.0068 | 0.0155 | 0.0145 | 0.0120 | 0.0109 |
| | $\mu_{LR}$ | 0.058 | 0.120 | 0.271 | 0.834 | 0.782 | 0.943 |
| | $M_{atom} = \sigma_{LR}^2$ | 0.0533 | 0.1093 | 0.1838 | 0.0994 | 0.1349 | 0.0532 |
| | $\mu_S$ | 0.980 | 0.952 | 0.882 | 0.625 | 0.649 | 0.561 |
| | $\sigma_S^2$ | 0.0081 | 0.0198 | 0.0392 | 0.0334 | 0.0406 | 0.0235 |
| | $M_{mix} = \mu_S \cdot \sigma_S^2$ | 0.0079 | 0.0187 | 0.0333 | 0.0214 | 0.0270 | 0.0132 |
| $g = 0.5$ $d = 0.5$ | $\mu_I$ | 0.0009 | 0.0011 | 0.0015 | 0.0026 | 0.0013 | 0.0104 |
| | $\sigma_I^2$ | 3.72E-05 | 3.14E-05 | 3.69E-05 | 5.97E-05 | 2.55E-05 | 9.25E-05 |
| | $M_{mob} = \mu_L$ | 0.005 | 0.012 | 0.021 | 0.039 | 0.029 | 0.269 |
| | $\sigma_L^2$ | 0.0010 | 0.0028 | 0.0049 | 0.0091 | 0.0065 | 0.0068 |
| | $\mu_{LR}$ | 0.034 | 0.063 | 0.098 | 0.164 | 0.133 | 1.011 |
| | $M_{atom} = \sigma_{LR}^2$ | 0.0378 | 0.0712 | 0.1086 | 0.1589 | 0.1366 | 0.0903 |
| | $\mu_S$ | 0.990 | 0.974 | 0.956 | 0.921 | 0.935 | 0.483 |
| | $\sigma_S^2$ | 0.0033 | 0.0131 | 0.0242 | 0.0393 | 0.0329 | 0.0433 |
| | $M_{mix} = \mu_S \cdot \sigma_S^2$ | 0.0032 | 0.0125 | 0.0221 | 0.0357 | 0.0298 | 0.0211 |

In Teal organizations, although there is no central leader and roles (communities) are naturally differentiated, we expect a balance between communication within communities (similarity) and between communities (dissimilarity), as in mixbiotic societies. Therefore, Fig 4 shows a comparison of the calculation results of the mixism measure out of the results shown in Table 4.

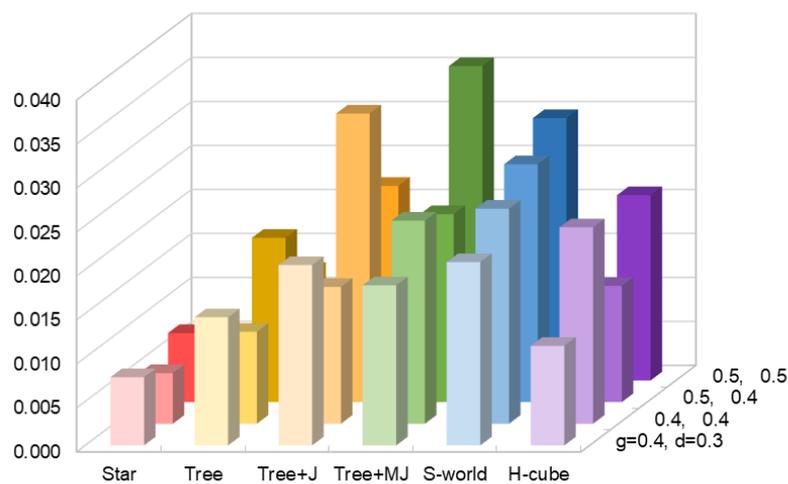

**Fig 4. Comparison of calculation results for mixism measure.**

Fig 4 shows that the values for tree+jumpers (Orange), tree+more jumpers (Green), and small-world (Teal) are larger than those for star (Red) and tree (Amber). This indicates that in Red and Amber organizations, communication is vertical to the central leader, whereas in Orange, Green, and Teal organizations, communication is horizontal among members. Comparing Orange, Green, and Teal organizations, although there is ruggedness in four cases, on average, Teal organization has the highest values of the mixism measure. This indicates that Teal organization balances similarity and dissimilarity in communication, and is living and mixbiotic between order and chaos. The lower value of the mixism measure for the hypercube (Flat) compared to Teal organization is inferred to be due to the fact that all members are equal and therefore less similar, as indicated by $\mu_S$ values in Table 4, i.e., communication is more distributed.

Next, Fig 5 shows two radar charts comparing the values of nine measures for six networks, in the case of $g = 0.4, d = 0.4$ representing four cases. In this chart, for ease of viewing, the values of each item are normalized using maximum value. Fig 5A is drawn for six networks and Fig 5B for five networks excluding the hypercube.

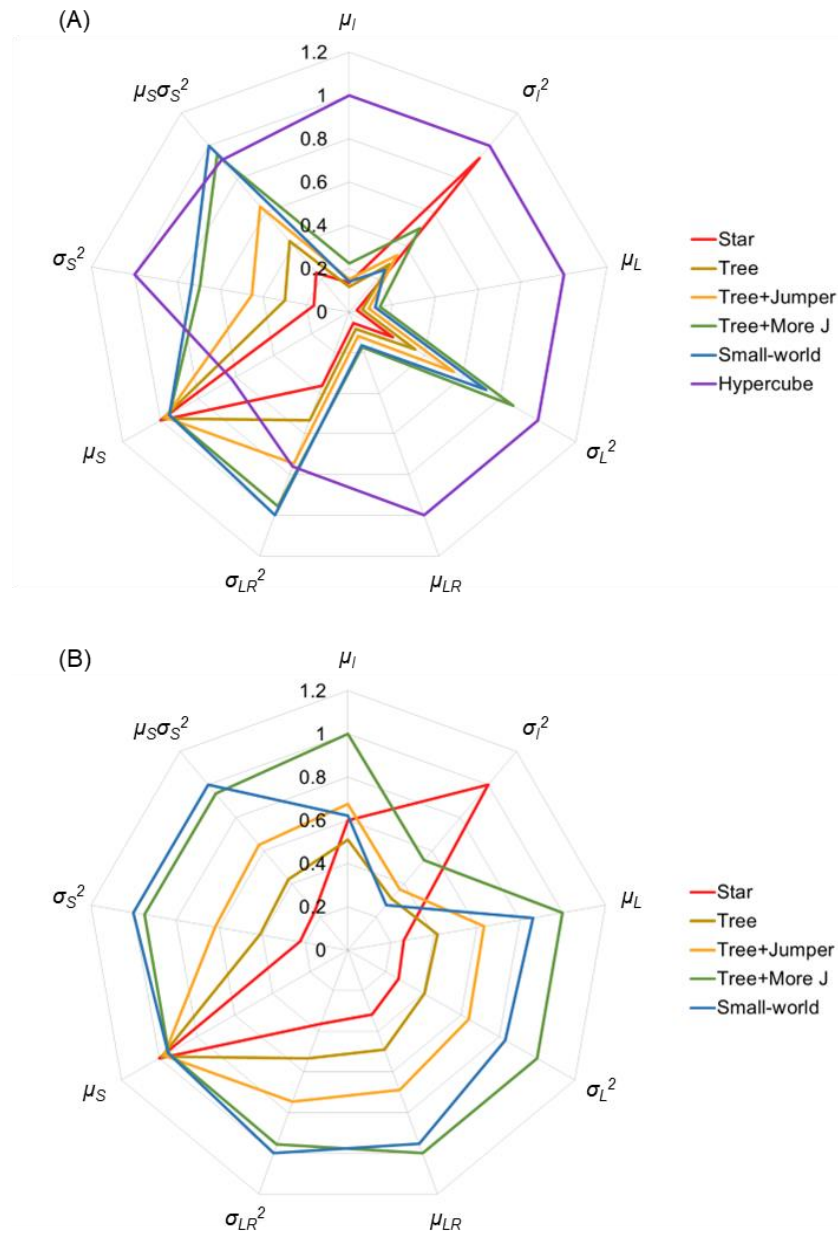

**Fig 5. Rader chart of calculation results for mixbiotic society measures.**

Fig 5A shows that the hypercube (Flat) is characterized by large values for all indicators except $\mu_S$. The large values of $\mu_I$, $\sigma_I^2$, $\mu_L$, $\sigma_L^2$, etc. indicate that information changes rapidly. As mentioned in the description of Fig 4, the similarity of communication is small and distributed because all members are equal. However, in a real organization, it is unlikely to be perfectly even and flat.

In Fig 5B, the star (Red) is somewhat unique, but the values of measures increase as one goes from the tree (Amber) through the tree+jumpers (Orange) and tree+more jumpers (Green) to the small-world (Teal). The fact that Teal organization has smaller values of $\mu_I$ and $\sigma_I^2$ than Green organization is because information is not concentrated and accumulated in the central leader, which indicates that the organization is decentralized. Therefore, considering that Teal organization has the largest value of mixism measure $\mu_S \cdot \sigma_S^2$, Teal organization is considered to be the most desirable organizational structure.

Fig 6 shows the trajectories of communication for the case $g = 0.4, d = 0.3$, drawn following the literature [10]. In the polar coordinate system, the magnitude of the $n$-dimensional vector of the information set $Q(t)$ is taken as the dynamic radius $r$, the angle formed by the $n$-dimensional vector and the unit vector $\mathbf{1}$ is taken as the declination angle $\theta$, and the trajectory is drawn by connecting the points at each time from $t = 0$ to $t_{max}$ in sequence.

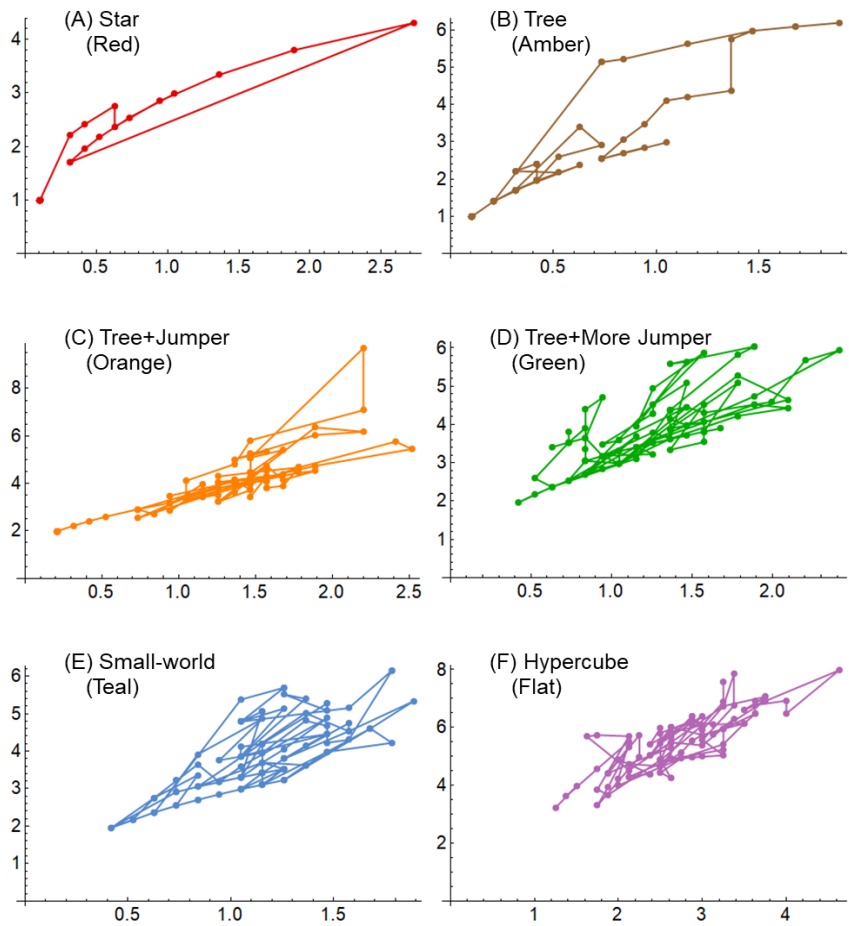

**Fig 6. Communication trajectories.**

The star in Fig 6A (Red) and the tree in Fig 6B (Amber) show that the trajectories are simple and unidirectional. This is because in Red and Amber organizations, communication is directed only toward the central leader. We can see that the communication trajectories widen as moving from the tree+jumpers in Fig 6C to the tree+more jumpers in Fig 6D or the small-world (Teal) in Fig 6E. The hypercube (Flat) in Fig 6F has a larger horizontal value and an even wider trajectory than the small-world (Teal) in Fig 6E. The wide trajectory of Teal organization means that the vector magnitude can change (the information amount can be larger or smaller) and the declination angle can change (communication can take place among various members). In other words, except for impractical Flat organization, Teal organization is considered the most desirable.

Note that Table 4 and Figs 4–6 are the results of random generation and disappearance of communication according to the computational flow shown in Fig 1. Since a real society is not random and there is a communication bias depending on an organizational structure, the results are likely to show more differences in practice. In other words, even in random simulations, MSM reflects well the differences in organizational structure.

## Discussion

Evaluation by MSM for communication networks corresponding to five typologies of organizational structure (Red, Amber, Orange, Green, and Teal) showed that Teal organization is desirable as an organizational structure, like a living organization on the edge of chaos. In particular, the higher value of Teal organization's mixism measure suggests that appropriate communication occurs within community (similarity) and between communities (dissimilarity) in organization, depending on the division of roles.

Teal organization is one form of mixbiotic societies in workplace and in corporate collaboration. In other words, evaluating an organization using MSM can be an assessment for organizational change. For example, an MSM assessment for actual communication data, such as those presented in the literature [10], and a correlation analysis with measurements of organizational efficiency and performance, as well as members' psychological and physical activity, would provide useful consulting for the goal achievement and persistent evolution in organization.

Teal organizations are often cited for organizational transformation in companies, but in the future, digital democratic organizations and platform cooperatives will also be targeted. By collecting real-time communication data and evaluating MSM on these information platforms, it will be possible to operate in a way that balances freedom and solidarity. Although MSM is still a new measure, it is expected to be effective in various fields and to be verified and improved through field data, which both will lead to the development of mixbiotic-society organizations.

# Acknowledgements

The authors received valuable comments on the perspectives of this study from the Hitachi Kyoto University Laboratory of the Kyoto University Open Innovation Institute. The authors would like to express their deepest gratitude. This work was supported by JST RISTEX Grant Number JPMJRS22J5, Japan.